\documentclass[12pt]{article} 
\title{ Rotations and $e$, $\nu$ Propagators, Part II}  
\author{{\it Richard Shurtleff~}\thanks{affiliation and mailing 
address: Department of Applied Mathematics and Sciences, 
Wentworth Institute of Technology, 550 Huntington Avenue, 
Boston, MA, USA, ZIP 02115, telephone number: (617) 989-4338, fax 
number: (617) 989-4591 , e-mail address: shurtleffr@wit.edu}} 
\begin{document} 
          
\maketitle               
			\begin{abstract}  

We continue to derive spacetime quantities and spin 1/2 propagators from rotations. Rotation-invariant projection operators are found for each element of a four element basis, i.e. a basis for four component quantities with specific transformation rules under rotations. With these four projection operators, we make two spacetime invariant projection operators, i.e. once space, time, energy, and momentum are identified. The spacetime invariant operators are propagators for free neutrinos. Except for the substitute basis, the process is the same as the one that gave electron propagators in Part I. 

PACS number(s): 11.30.-j, 11.30.Cp, and 03.65.Fd 
 
			\end{abstract}
\pagebreak

\section{Introduction} \label{intro} % 1 

	It is necessarily true that spacetime quantities can be obtained from rotation considerations because rotations form a subgroup of the group of spacetime transformations. The overall process \cite{partI} is to (i) define suitable rotation-founded quantities, i.e. 2-vectors and rotation matrices, etc, (ii) specify a basis for pairs of 2-vectors in Sec.~2, (iii) obtain rotation invariant projection operators for the basis in Sec.~3, and (iv) in Sec.~4 identify energy and momentum as functions of the parameters of the basis and find spacetime symmetric combinations of the projection operators.

	Energy-momentum appears both as a factor with space and time in a phase angle and it appears as a factor with gamma matrices in a matrix composed of outer products of basis pairs. For the phase, a wide range of functions would suffice, but the outer products give specific functions of the basis parameters. Therefore, by changing the basis from Part I, we generate new outer products and obtain different energy-momentum functions. 

	The energy-momentum functions have null magnitude and the spacetime invariant projection operators are the Green's functions for a relativistic spin 1/2 single particle wave equation. Thus the projection operators are propagators for a free neutrino or antineutrino.

%\pagebreak

\section{A Basis with Mixed Eigenvectors} \label{likeneu}% 2

The choice of the $uv$ basis pairs in Part I \cite{partI} started the process that lead to the electron propagator. So, to obtain a new propagator, we choose a different basis.

	In each of the $uv$ basis pairs $u_{+}^{+},$ $u_{-}^{-},$ $v_{+}^{+},$ and $v_{-}^{-},$ see I(9) in Part I, the upper and lower 2-vectors are proportional to the same eigenvector, either $u^{+}$ or $u^{-},$ where the eigenvectors are displayed in I(3). Consider now a new basis obtained by exchanging the spin up eigenvectors $u^{+}$ with the spin down eigenvectors $u^{-}$ in the lower 2-vectors. We get 
  $$	%notes, 2/18/0
 f_{-}^{+} = \pmatrix{ e^{w/2} u^{+} \cr e^{-w/2} u^{-} } \hspace{1in} f_{+}^{-} = \pmatrix{ e^{ -w/2} u^{-} \cr e^{w/2} u^{+} }    ,$$ 
\begin{equation} 	\label{fg} %kernel5, October99 tex file
 g_{-}^{+} = \pmatrix{  e^{-w/2} u^{+} \cr  - e^{w/2} u^{-} } \hspace{1in} g_{+}^{-} = \pmatrix{  - e^{ w/2} u^{-} \cr  e^{-w/2} u^{+} }    .
\end{equation}

	We rotate the 2-vectors so that the pairs acquire a common phase, \begin{equation}	\label{Rfg}
R_{-}^{+} f_{-}^{+} =  e^{+i\theta/2} f_{-}^{+} \hspace{0.5cm} R_{+}^{-} f_{+}^{-} =  e^{+i\theta/2} f_{+}^{-} \hspace{0.5cm} R_{-}^{+} g_{-}^{+} =  e^{+i\theta/2} g_{-}^{+} \hspace{0.5cm} R_{+}^{-} g_{+}^{-} =  e^{+i\theta/2} g_{+}^{-}  ,
\end{equation}
where $R_{-}^{+}$ means applying $R$ to the upper 2-vector and $R^{-1}$ to the lower 2-vector. One can show that the rotated basis pairs (\ref{Rfg}) are normalized to $2 \cosh{w}$ and they are mutually orthogonal, 
\begin{equation}	\label{norm,fg} %notes 2/19/0
 i^{\dagger} i = 2 \cosh w \hspace{1cm} i^{\dagger} j = 0, \hspace{1cm} i,j \in \{R_{-}^{+} f_{-}^{+}, R_{+}^{-} f_{+}^{-},   R_{-}^{+} g_{-}^{+}, R_{+}^{-} g_{+}^{-} \},
\end{equation}
where there is no sum over $i$ in the left expression and $i$ is not the same as $j$ in the middle expression.

 \section{Projection Operators for the Mixed Basis} \label{projections} % 4

	The process of obtaining the projection operators for the $fg$ basis goes through just as with the $uv$ basis in Sec.~I4. Any four component object $\psi$ can be expressed in the $Rfg$ basis (\ref{Rfg}),
\begin{equation}	\label{psi0}
\psi = e^{+i\theta/2} \psi_{0} = e^{+i\theta/2}(a f_{-}^{+} + b f_{+}^{-} + c g_{-}^{+} + d g_{+}^{-}), 
\end{equation}
where $a$ = ${f^{+}_{-}}^{\dagger} \psi_{0}/ (2 \cosh{w})$ and $\psi_{0}$ is $\psi$ for $\theta$ = 0. We need to obtain four projection operators, each operator selecting the part of $\psi$ that is proportional to one of the basis pairs. We split the phase into separate terms, thereby defining $\Delta,$
\begin{equation}	\label{psi,theta}
\frac{\theta}{2} = - \Delta + p^{k} x^{k},
\end{equation}
where the $p^{k}$ are three as-yet-unidentified functions of the ratio parameter $w$ and the unit vector $n^{k}.$ Together with the $x^{k}$ the $p^{k}$ constitute the parameters of a three dimensional delta function for selecting $w$ and $n^{k}.$ $\Delta$ is the difference between the rotated basis phase $\theta/2$ and the delta function phase. For details see Sec.~I4.

 	Following the same steps that lead to I(22), the projection operator $K(2,1,R_{-}^{+} f_{-}^{+}) \gamma^{4},$ for $f_{-}^{+}$ is found to be  
\begin{equation}	\label{K(2,1)}
 K(2,1,R_{-}^{+} f_{-}^{+}) \gamma^{4} \equiv \int \frac{d^3 p^{\prime} }{(2 \pi)^3} \frac{1}{2 {\cosh{(w)}}^{\prime}}e^{i \theta_{2}^{\prime}/2}  f_{-}^{+ \, \prime} {f_{-}^{+ \, \prime }}^{\dagger} e^{ -i \theta_{1}^{\prime}/2} .
\end{equation} 
By straightforward calculation one can verify that
\begin{equation}	\label{K(2,1)psi}
 \int d^{3} x_{1} K(2,1,R_{-}^{+} f_{-}^{+}) \gamma^{4} \psi(1) =  e^{i \theta_{2} / 2} (a f_{-}^{+}) = \psi(2)_{b=c=d=0} ,
\end{equation} 
where $\psi(1)$ is $\psi$ with $\theta$ = $\theta_{1}$ and $ \theta_{i}/2 $ = $- \Delta_{i}^{\prime} +  {p^{k }}^{\prime} x_{i}^{k} .$
Comparing (\ref{K(2,1)psi}) with (\ref{psi0}), we see that the projection operator recovers the desired portion of $\psi$ with a new phase $\theta_{2} / 2.$ The projection operators for the other basis pairs differ from $K(2,1,R_{-}^{+} f_{-}^{+}) \gamma^{4}$ in the matrices by substituting $f_{+}^{-},$ $g_{-}^{+},$ and $g_{+}^{-}$ in place of $f_{-}^{+}$ in (\ref{K(2,1)}).

	The appearance of the scalar product $p^{k} x^{k}$ and the expansion of $f_{-}^{+ \, \prime} {f_{-}^{+ \, \prime }}^{\dagger}$ in (\ref{f-+f-+g}) and Problem 1 below shows that $K(2,1,R_{-}^{+} f_{-}^{+})\gamma^{4}$ is a rotation invariant projection operator and the rotation angle $\theta$ is also a rotation invariant when $\Delta$ and $\gamma^{4}$ are rotation invariants. We turn now to making the definitions and finding the combinations that give spacetime invariance.

\section{Space-Time Symmetry, the Neutrino Propagator} \label{space-time} % 3

	In this section it is found that the projection operators can have space-time symmetry, but only after combining two projection operators.  There are three parts to this, just as in Part I. The three dimensional integral over $x_{1}$ in (\ref{K(2,1)psi}) becomes a surface integral in four dimensional space-time. The phase difference $\Delta$ is shown to be the energy times the time, thereby making the rotation angle an invariant. The remaining factor is a non-invariant matrix for one projection operator, but two such matrices make a space-time invariant quantity for a particular choice of momentum function.

	{\textit{Time}}. Time can now be introduced by applying an integral expression for the phase factor $e^{-i\Delta}$, see I(25). 	By (\ref{K(2,1)}) and (\ref{K(2,1)psi}), we have
	$$\psi(2)_{b=c=d=0} = \int d^3x_{1}K(2,1,R_{-}^{+}f_{-}^{+}) \gamma^{4} \psi(1) $$ $$ =  \int d^3x_{1}\int \frac{d^3 p^{\prime} }{(2 \pi)^3} \frac{1}{2 {\cosh{(w)}}^{\prime}}e^{i \theta_{2}^{\prime}/2}  f_{-}^{+ \, \prime} {f_{-}^{+ \, \prime }}^{\dagger} e^{ -i \theta_{1}^{\prime}/2}e^{i \theta_{1} / 2} \gamma^{4} \gamma^{4} \psi_{0}
$$
\begin{equation}	\label{K-+K-+psi}
= \int d^3 x_{1}\lbrack \int \frac{d^3 p^{\prime} }{(2 \pi)^3} \frac{1}{2 {\cosh{(w)}}^{\prime}}e^{i \theta_{2}^{\prime}/2}  f_{-}^{+ \, \prime} {f_{-}^{+ \, \prime }}^{\dagger} \gamma^{4}\rbrack e^{ i (\Delta_{1}^{\prime}- \Delta_{1})} e^{i (p^{k \; \prime} - p^{k})x_{1}^{k}}  \gamma^{4} \psi_{0}.
\end{equation}
Let $I$ be the quantity in (\ref{K-+K-+psi}) in brackets. By (\ref{psi,theta}) and I(24) and for $\Delta_{2}^{\prime} >$ 0, we get
$$ I = \int \frac{d^3 p^{\prime} }{(2 \pi)^3} \frac{1}{2 {\cosh{(w)}}^{\prime}} e^{-i\Delta_{2}^{\prime}} e^{ip^{k \; \prime}x_{2}^{k}} f_{-}^{+ \, \prime} {f_{-}^{+ \, \prime }}^{\dagger} \gamma^{4} 
$$
$$= \int \frac{d^3 p^{\prime} }{(2 \pi)^3} \frac{1}{2 {\cosh{(w)}}^{\prime}}  ( \frac{i}{ \pi}  \int_{-\infty}^{\infty} da \frac{e^{-ia\Delta_{2}^{\prime}}}{a^2-1+i\epsilon}) e^{ip^{k \; \prime}x_{2}^{k}} f_{-}^{+ \, \prime} {f_{-}^{+ \, \prime }}^{\dagger} \gamma^{4} 
$$
\begin{equation} \label{QED}
=  i  \int \frac{d^4 p^{\prime} }{(2 \pi)^4} e^{-i(p^{4 \; \prime}t_{2} - p^{k \; \prime}x_{2}^{k})}  \frac{ (\pm m) f_{-}^{+ \, \prime} {f_{-}^{+ \, \prime }}^{\dagger} \gamma^{4}}{{p^{4 \; \prime}}^2- m^{2} \cosh^{2}{w}^{\prime} + i\epsilon} 
\end{equation}
where we introduce a fourth component of momentum and a fourth component of displacement, i.e. energy and time,
\begin{equation}	\label{p4t}
{p^{4}}^{\prime} \equiv \pm m {\cosh{(w)}}^{\prime} a   \hspace{1cm}  a \Delta_{2}^{\prime} = {p^{4}}^{\prime} t_{2} , 
\end{equation}
where the sign is to be determined below and $m$ is the `mass,' a term to be discussed below.

	{\textit{Surface Integral}}. Next, in (\ref{K-+K-+psi}) identify the integral over the 3-space $x_{1}^{k}$ with a surface integral in space-time,
\begin{equation}	\label{K-+K-+psi1}
\psi(2)_{b=c=d=0} = \int_{S} d^4 x_{1}I e^{ i (\Delta_{1}^{\prime}- \Delta_{1})} e^{i (p^{k \; \prime} - p^{k})x_{1}^{k}} N_{\mu} \gamma^{\mu} \psi_{0}
\end{equation}
where $x^{4}$ = $t,$ $\mu \in$ $\{1,2,3,4\},$ $N^{\mu}$ = $\{N^{k},N^{4}\}$ is the unit normal to the three dimensional surface of integration $S$ in four dimensional space-time, and the space-time summation convention is used, $N_{\mu} \gamma^{\mu}$ = $N^{4} \gamma^{4} - N^{k} \gamma^{k}.$ In (\ref{K(2,1)psi}) and (\ref{K-+K-+psi}), the integration is over the three dimensional surface $x_{1}^{k}$ in the special space-time reference frame with the normal in the time direction,  $N^{\mu}$ = $\{0,0,0,1\}.$ 
 
	{\textit{Matrices}}. The projection operator $K(2,1,R_{-}^{+}f_{-}^{+}) \gamma^{4}$ rewritten with (\ref{K-+K-+psi}) and (\ref{QED}) fails to have space-time symmetry because of the quantities $m {f_{-}^{+}}^{\prime} {{f_{-}^{+}}^{\prime}}^{\dagger}  \gamma^{4}$ and ${p^{4 \; \prime}}^2-{(m {\cosh{w}}^{\prime})}^2 $ in (\ref{QED}). We can express the matrix $m {f_{-}^{+}}^{\prime} {{f_{-}^{+}}^{\prime}}^{\dagger}  \gamma^{4}$ as a sum of sixteen linearly independent $4 \times 4$ matrices such as the set of gamma matrices in Part I, Appendix A. By (\ref{fg}) and I(49), we get 
$$ m {f_{-}^{+}} {{f_{-}^{+}}}^{\dagger}  \gamma^{4} = \frac{1}{2}[ m {\cosh{(w)}} \gamma^{4} - m {\cosh{(w)}}n^{j} \gamma^{j} + i {\sinh{(w)}} n^{k} \gamma^{k} \gamma^{5} - i {\sinh{(w)}} \gamma^{4} \gamma^{5} ] + 
$$
\begin{equation}	\label{f-+f-+g}
+\pmatrix{u^{+} {u^{-}}^{\dagger} && 0 \cr 0 && u^{-} {u^{+}}^{\dagger} },
\end{equation}
where we drop the primes and the last term on the right is messy when written in terms of the $\gamma^{A}$s. It can be shown that the matrix is not a space-time invariant when the four gammas $\gamma^{\mu}$  in (\ref{f-+f-+g}) transform as a space-time 4-vector. 

	Similarly one finds the expansion of $ m {g_{-}^{+}}^{\prime} {{g_{-}^{+}}^{\prime}}^{\dagger}  \gamma^{4}$ over the same set of gammas. We have
$$m {g_{-}^{+}} {{g_{-}^{+}}}^{\dagger}  \gamma^{4} = \frac{1}{2}[ m {\cosh{(w)}} \gamma^{4} - m {\cosh{(w)}}n^{j} \gamma^{j} - i {\sinh{(w)}} n^{k} \gamma^{k} \gamma^{5} + i {\sinh{(w)}} \gamma^{4} \gamma^{5} ] + 
$$
\begin{equation}	\label{g-+g-+g}
-\pmatrix{u^{+} {u^{-}}^{\dagger} && 0 \cr 0 && u^{-} {u^{+}}^{\dagger} },
\end{equation}
When we add the two expressions, we get
\begin{equation}	\label{Epmatrix1}
m (  {f_{-}^{+}} {{f_{-}^{+}}}^{\dagger}  + {g_{-}^{+}} {{g_{-}^{+}}}^{\dagger} ) \gamma^{4} = m {\cosh{(w)}} \gamma^{4} - m {\cosh{(w)}}n^{k} \gamma^{k} .
\end{equation}
Let $\gamma^{k}$ and $\gamma^{4}$ form a 4-vector, i.e. transform as in I(28).  If $m \cosh{(w)} n^{k}$ and $m \cosh{w}$ also transform as the components of a 4-vector, then the expression (\ref{Epmatrix1}) is a space-time invariant. 

	{\textit{Energy-momentum}}. Energy and momentum have a special role because the energy-momentum 4-vector occurs (i) with $x^{\mu}$ in the invariant delta function phase $p_{\mu} x^{\mu}$ and also (ii) with the gamma matrices in the invariant polarization matrices $p_{\mu} \gamma^{\mu}.$ Hence, as in Part I, we define the momentum and energy functions by comparing components. We get
\begin{equation} \label{pk}
p^{k} = m \cosh{(w)} n^{k} \hspace{1cm} p^{4} = + m\cosh{(w)} a .
\end{equation}
Then the pole (its at $a$ = 1) in (\ref{QED}) occurs when $p^{4}$ is the energy $E,$
\begin{equation}	\label{E}
E \equiv + m \cosh{w} \hspace{1cm} \Delta = E t.
\end{equation} 

	Unlike the $m$ in Part I for an electron where $m$ is the magnitude of the electron energy-momentum space-time four vector, the quantity $m$ may not be made an invariant as easily. The difficulty is that the smallest $\cosh{w}$ can be is 1 while the energy $E$ can be made as small as desired by a suitable red-shifting boost. See Problem 3.

	$E > 0$ {\textit{Propagator}}. Combining (\ref{K-+K-+psi}), (\ref{QED}), (\ref{Epmatrix1}), and (\ref{pk}) gives an expression that is a space-time invariant. We have 
\begin{equation}	\label{QEDnu1}
\psi(2)_{b=d=0} = i\int \frac{d^4 x_{1} d^4 p^{\prime} }{(2 \pi)^4} e^{-ip_{\mu}^{\prime}x_{2}^{\mu}} \; \frac{ {p_{\nu}}^{\prime} \gamma^{\nu} }{p_{\tau}^{\prime}p^{\tau \, \prime}- i\epsilon} e^{ip_{\eta}^{\prime}x_{1}^{\eta}}  N_{\sigma} \gamma^{\sigma} \psi(1).
\end{equation}

	$E > 0$ {\textit{Propagator}}. The projection operators for the remaining two $fg$ basis pairs gives
\begin{equation}	\label{Epmatrix2}
- m (  {f_{+}^{-}}^{\prime} {{f_{+}^{-}}^{\prime}}^{\dagger}  + {g_{+}^{-}}^{\prime} {{g_{+}^{-}}^{\prime}}^{\dagger} ) \gamma^{4} = -m {\cosh{(w)}}^{\prime} \gamma^{4} - m {\cosh{(w)}}^{\prime}n^{k} \gamma^{k} .
\end{equation}
The new matrix gives the same momentum function but negative the energy function in (\ref{pk}). Hence we now define the energy-momentum functions to be
\begin{equation} \label{pk2}
p^{k} \equiv m \cosh{(w)} n^{k} \hspace{1cm} p^{4} \equiv - m\cosh{(w)} a .
\end{equation}
Then the pole (its at $a$ = 1) in (\ref{QEDnu1}) occurs when $p^{4}$ is the negative energy $E,$
\begin{equation}	\label{E2}
E \equiv - m \cosh{w} \hspace{1cm} \Delta = E t.
\end{equation}

	Note that (\ref{pk}) and (\ref{pk2}) differ by the sign of $m\cosh{w}$ for $p^{4}.$ This is important because the wave function $\psi$ which is originally given, by (\ref{fg}), (\ref{Rfg}), and (\ref{psi0}), as a function of the rotation angle $\theta,$ ratio parameter $w,$ and unit 3-vector $n^{k},$ is rewritten in terms of $p^{\mu}$ so that the delta functions can be applied. That step is necessary because the delta functions are in terms of $p^{\mu},$ see Sec.~I4.

	The sum of the projection operators for $f_{+}^{-}$ and $g_{+}^{-}$ can be written as
\begin{equation}	\label{QEDnu2}
\psi(2)_{a=c=0} = i\int \frac{d^4 x_{1} d^4 p^{\prime} }{(2 \pi)^4} e^{-ip_{\mu}^{\prime}x_{2}^{\mu}} \; \frac{ {p_{\nu}}^{\prime} \gamma^{\nu}}{p_{\tau}^{\prime}p^{\tau \, \prime}- i\epsilon} e^{ip_{\eta}^{\prime}x_{1}^{\eta}}  N_{\sigma} \gamma^{\sigma} \psi(1).
\end{equation}

	{\textit{Experimental confirmation}}. The fractions in (\ref{QEDnu1}) and (\ref{QEDnu2}) are sometimes written as inverse matrices, $({p_{\mu}}^{\prime} \gamma^{\mu})^{-1}$ because
\begin{equation}	\label{m-1,again}
({p_{\mu}}^{\prime} \gamma^{\mu})({p_{\nu}}^{\prime} \gamma^{\nu}) = {p^{4 \; \prime}}^{2} - {p^{k \; \prime}}^2  = p_{\tau}^{\prime}p^{\tau \, \prime}
\end{equation}
On this basis the propagator for positive energy (\ref{QEDnu1}) and the propagator for negative energy (\ref{QEDnu2}) can be shown to be the Green's functions for the matrix operator $({p_{\mu}}^{\prime} \gamma^{\mu}).$ In the representation of the gammas here, the upper and lower 2-vectors become separated in the wave equations obtained for the Green's functions. Such wave functions conventionally describe free neutrinos. (The problem of whether or not all four components of a neutrino take part in interactions is immaterial for the `free' neutrino.) We conclude that the propagators (\ref{QEDnu1}) and (\ref{QEDnu2}) can describe the free neutrino and antineutrino.

	{\textit{Time, energy, and rotation angle}}. By (\ref{psi,theta}), (\ref{E}) and (\ref{E2}) and for both positive and negative energy, we have
\begin{equation}	\label{t+-}
t = \frac{1}{2E} (2p^{k}x^{k} - \theta) 
\end{equation}
We make $t_{1}$ = 0 on the surface of integration $x_{1}^{k}.$ Next we arrange that $t_{2}$ = 0 on the surface $x_{2}^{k}$ = $x_{1}^{k}.$ By (\ref{t+-}), at any point $x^{k},$
$$
{\mathrm{positive}} \hspace{0.25 cm} {\mathrm{energy}} \hspace{0.3 cm} E > 0 \hspace{0.3 cm} {\mathrm{and}} \hspace{0.3 cm} \theta \hspace{0.15 cm} {\mathrm{decreasing}} \hspace{0.3cm} \Rightarrow \hspace{0.3cm}  t_{2} > 0  \hspace{0.3 cm}	{\mathrm{and}} \hspace{0.3 cm} t_{2} \hspace{0.15 cm} {\mathrm{increasing}} \hspace{1cm}
$$
\begin{equation}	\label{t+-a}
{\mathrm{negative}} \hspace{0.25 cm} {\mathrm{energy}} \hspace{0.3 cm} E < 0 \hspace{0.3 cm} {\mathrm{and}} \hspace{0.3 cm} \theta \hspace{0.15 cm} {\mathrm{decreasing}} \hspace{0.3cm} \Rightarrow \hspace{0.3cm}  t_{2} < 0  \hspace{0.3 cm}	{\mathrm{and}} \hspace{0.3 cm} t_{2} \hspace{0.15 cm} {\mathrm{decreasing}} .
\end{equation}
The primes have been dropped. As we found in Part I for the electron, we find here for the neutrino that the rotation angle $\theta$ decreases as the particle moves in time away from the surface $x^{k}$ at $t$ = 0. 

\pagebreak
	
\appendix

%\pagebreak

 \section{Problems} % A

\noindent 1. (a) Derive (\ref{f-+f-+g}), (\ref{g-+g-+g}), and (\ref{Epmatrix1}). (b) Express the matrix displayed in (\ref{f-+f-+g}) and (\ref{g-+g-+g}) linearly in terms of the gammas, I(47). (c) Show that the matrices (\ref{f-+f-+g}) and (\ref{g-+g-+g}) are rotation invariants.

\vspace{0.3cm}
\noindent 2. For a massive particle, the proper time $\tau$ along a path $x^{\mu}$ that is proportional to $p^{\mu}$ is the ordinary time $t$ in a space-time reference frame with $p^{k}$ = 0. How is $\tau$ related to the rotation angle $\theta?$ (b) For the neutrino, find straight line paths $x^{\mu}$ in space-time with $x^{k}$ proportional to $p^{k}$ and that satisfy (\ref{t+-}). How does the proper time on these paths depend on the internal spin space rotation angle $\theta?$ (See \cite{backnu})

\vspace{0.3cm}
\noindent 3. Suppose the parameter $m$ in the expressions (\ref{E}) and (\ref{E2}) for energy $E$ is a constant for neutrinos independent of reference frame. Consider a source of neutrinos each with energy $E > m > 0$ and momentum $p^{k}$ = $\{E,0,0\},$ both measured in the rest frame of the source. (a) Find the velocity of the frame with minimum energy $E_{0}$ = $m$ and $p^{k}$ = $\{m,0,0\}.$ (b) Show that boosts from the frame in (a) along $+x$ are not allowed. (c) Free neutrinos would move at the speed of light in the frame in (a) (at least on average over long distances). Would neutrino scattering be affected? If so, how?


\begin{thebibliography}{9}

\bibitem{partI} Shurtleff, R., hep-th/0007232, and references therein.

\bibitem{backnu} Shurtleff, R., hep-ph/0003071.


\end{thebibliography}
\end{document}